\documentstyle[psfig,11pt,aaspp4]{article}   

\slugcomment{To appear in The Astrophysical Journal}
\lefthead{Pastoriza et al.}
\righthead{The central 2 kpc of NGC\,3379}

\begin{document}

\newcommand{\dapp}{$\approx$}
\newcommand{\app}{$\sim$}
\newcommand{\prop}{$\propto$}
\newcommand{\tento}[1]{$10^{#1}$}
\newcommand{\vzs}{$\times$}
\newcommand{\lb}{$\lambda$}
\newcommand{\mm}{$\pm$}
\newcommand{\smle}{$\ddot{\smile}$}
\newcommand{\nd}{\nodata}
\newcommand{\lsol}{$L_\odot$}
\newcommand{\msol}{$M_\odot$}
\newcommand{\ebv}{\hbox{$E(B\!-\!V)$}}
\newcommand{\hbc}{H$_\circ$}
\newcommand{\lbol}{$L_{bol}$}
\newcommand{\mdot}{$\dot{M}$}
\newcommand{\acr}{\msol\,yr$^{-1}$}
\newcommand{\nel}{N$_e$}
\newcommand{\tel}{T$_e$}
\newcommand{\teo}{T$_e$\,[\ion{O}{3}]}
\newcommand{\jmh}{\hbox{$(J\!-\!H)$}}
\newcommand{\hmk}{\hbox{$(H\!-\!K)$}}
\newcommand{\kms}{km\,s$^{-1}$}
\newcommand{\ergs}{ergs\,s$^{-1}$}
\newcommand{\ergcs}{ergs\,cm$^{-2}$\,s$^{-1}$}
\newcommand{\ergcsa}{ergs\,cm$^{-2}$\,s$^{-1}$\,$\AA^{-1}$}
\newcommand{\cmc}{cm$^{-3}$}
\newcommand{\mic}{$\mu$m}
\newcommand{\caiik}{\ion{Ca}{2}\,K\,\lb3933}
\newcommand{\cn}{CN\,\lb4200}
\newcommand{\gband}{CH\,Gband\,\lb4301}
\newcommand{\mgi}{\ion{Mg}{1}}
\newcommand{\mgb}{\ion{Mg}{1}+MgH\,\lb5175}
\newcommand{\brgama}{Br$\gamma$}
\newcommand{\pagama}{Pa$\gamma$}
\newcommand{\pabeta}{Pa$\beta$}
\newcommand{\palfa}{Pa$\alpha$}
\newcommand{\halfa}{H$\alpha$}
\newcommand{\hbeta}{H$\beta$}
\newcommand{\hgama}{H$\gamma$}
\newcommand{\hdelta}{H$\delta$}
\newcommand{\heps}{H$\epsilon$}
\newcommand{\lya}{Ly$\alpha$}
\newcommand{\heii}{\ion{He}{2}\,\lb4686}
\newcommand{\hen}{\ion{He}{1}\,\lb5876}
\newcommand{\feii}{\ion{Fe}{2}}
\newcommand{\nif}{[\ion{N}{1}]\,\lb5199}
\newcommand{\niid}{[\ion{N}{2}]\,\lb6548,6584}
\newcommand{\oif}{[\ion{O}{1}]\,\lb6300}
\newcommand{\oii}{[\ion{O}{2}]\,\lb3727}
\newcommand{\oiiia}{[\ion{O}{3}]\,\lb5007}
\newcommand{\oiiib}{[\ion{O}{3}]\,\lb4363}
\newcommand{\oiiid}{[\ion{O}{3}]\,\lb4959,5007}
\newcommand{\siia}{[\ion{S}{2}]\,\lb6717,6732}
\newcommand{\siib}{[\ion{S}{2}]\,\lb4075}
\newcommand{\neiiia}{[\ion{Ne}{3}]\,\lb3869}
\newcommand{\neva}{[\ion{Ne}{5}]\,\lb34250}
\newcommand{\nevb}{[\ion{Ne}{5}]\,\lb3346}
\newcommand{\oit}{[\ion{O}{1}]\,\lb}
\newcommand{\oiiit}{[\ion{O}{3}]\,\lb}
\newcommand{\neiiit}{[\ion{Ne}{3}]\,\lb}
\newcommand{\niit}{[\ion{N}{2}]\,\lb}
\newcommand{\siit}{[\ion{S}{2}]\,\lb}
\newcommand{\heuv}{\ion{He}{2}\,\lb1640 + \ion{O}{3}]\,\lb1663}
\newcommand{\civ}{\ion{C}{4}\,\lb1549}
\newcommand{\hh}{H$_2$}
\newcommand{\bsiiia}{[\ion{S}{3}\,\lb9532}
\newcommand{\bsiiib}{[\ion{S}{3}\,\lb9068}
\newcommand{\bhen}{\ion{He}{1}\,1.083\mic}
\newcommand{\bfeiia}{[\ion{Fe}{2}]\,1.26\mic}
\newcommand{\bfeiib}{[\ion{Fe}{2}]\,1.66\,mic}
\newcommand{\bhepg}{\ion{He}{1}$+$\pagama}
\newcommand{\bhepd}{\ion{He}{2}$+$Pa$\delta$}
\newcommand{\nttsn}{NGC\,3379}
\newcommand{\vmi}{\hbox{$(V\!-\!I)$}}
\newcommand{\fcf}{$c_4$}


\title{A photometric and kinematic study of the stars and interstellar
medium  in the central two kpc of \nttsn\,\footnote{Based on
observations obtained at the ESO 3.6m telescope}}

\author{Miriani G. Pastoriza\altaffilmark{2,3}, 
	Cl\'audia Winge\altaffilmark{2,4},
 	Fabricio Ferrari\altaffilmark{2,4}
	F. Duccio Macchetto\altaffilmark{5,6}} 
\and 
\author{Nicola Caon\altaffilmark{7}}

\altaffiltext{2}{Instituto de F\'{\i}sica -- UFRGS, Av. Bento Gon\c{c}alves, 
9500, C.P. 15051, CEP 91501-950, Porto Alegre, RS, Brazil}
\altaffiltext{3}{e-mail: mgp@if.ufrgs.br}
\altaffiltext{4}{CNPq Fellowship}
\altaffiltext{5}{Space Telescope Science Institute, 3700 San Martin Drive,
Baltimore, MD21218, USA.} 
\altaffiltext{6}{Affiliated to the Astrophysics Division of ESA}
\altaffiltext{7}{Instituto de Astrofisica de Canarias, La Laguna, Tenerife,
Spain}

\begin{abstract}

HST images of \nttsn\ show that the V and I luminosity profiles in the
inner 13\arcsec\ of this E1 galaxy are represented by two different
components: a stellar bulge following a S\'ersic Law with exponent $n$
= 2.36, and a central core ($r < $ 0\farcs7) with a characteristic
``cuspy'' profile. Subtraction of the underlying stellar component
represented by the fitted S\'ersic profile revealed the presence of a
small ($r$ \dapp\ 105 pc) dust disk of about 150 \msol, oriented at PA
= 125\arcdeg\ and inclined \dapp\ 77\arcdeg\ with respect to the line
of sight. The same absorption structure is detected in the color-index
\vmi\ image. The stellar rotation in the inner 20\arcsec\ is well
represented by a parametric planar disk model, inclined
\dapp\ 26\arcdeg\ relative to the plane of the sky, and apparent major
axis along PA \dapp\ 67\arcdeg. The gas velocity curves in the inner
5\arcsec\ show a steep gradient, indicating that the gas rotates much
faster than the stars, although in the same direction. The velocity
field  of the gaseous system, however, is not consistent with the simple
model of Keplerian rotation sustained by the large (7 \vzs\ \tento{9}
\msol\ within a radius of \app\ 90 pc) central mass implied by the
maximum velocity observed, but the available data precludes a more
detailed analysis.

\end{abstract}


\section{Introduction}

High spatial resolution images obtained with the Hubble Space Telescope
revealed the presence of small nuclear stellar disks in several early
type galaxies (van den Bosch et al.\markcite{vdbetal94} 1994; van den
Bosch, Jaffe \& van der Marel\markcite{bjm98} 1998) as well as patchy
absorption due to dust near their nuclei (van Dokkum \&
Franx\markcite{vDf95} 1995). The presence of stellar disks in
elliptical galaxies has important consequences on the mechanisms of
galaxy formation since it suggests a continuity between spirals,
lenticulars, and ellipticals, implying that a single mechanism may
control the morphology of the protogalaxy (Bender et
al.\markcite{rbetal89} 1989; Capaccioli\markcite{c:m90} 1990). If, on
the other hand, it is assumed that elliptical galaxies are formed by
merger of spirals (Illingworth \& Franx\markcite{if89} 1989; Balcells
\& Quinn\markcite{bq90} 1990; Hernquist \& Barnes\markcite{hb91} 1991)
or coalescence of dwarf systems (Cot\'e, Marzke \& West\markcite{cmw98}
1998), it is difficult to explain the survival of such small disks over
large time scales after the merger event.

\nttsn\ is a bright E1 galaxy with almost perfectly elliptical
isophotes (Lauer\markcite{l:t95} 1985). Detailed surface photometry
based on photographic plates and CCD images covering the range from 18
mag arcsec$^{-2}$ to 29 mag arcsec$^{-2}$ yielded a mean ellipticity
of $\epsilon = 0.11$ and a major axis mean position of PA =
70\arcdeg\ (Capaccioli et al.\markcite{mcetal90} 1990). The isophote
twist is smaller than 5\arcdeg\ and the ellipticity varies by less than
0.06 between 10\arcsec\ and 100\arcsec\ (Capaccioli et
al.\markcite{mcetal90} 1990; Peletier et al.\markcite{rfpetal90}
1990).  The stellar rotation curves measured up to R = 30\arcsec\ show
evidence of slow (\app 50 \kms) rotation around the photometric minor
axis (Franx, Illingworth \& Heckman\markcite{fih89} 1989; Bender,
Saglia \& Gerhard\markcite{bsg94} 1994), and a mean stellar velocity
dispersion of 180 \kms. Higher velocity dispersion is observed in the
galaxy core ($r \lesssim$ 2\arcsec; Bender et al.\markcite{bsg94}
1994).

Although this galaxy is often regarded as the prototype for its
morphological type, there have been several suggestions that it may
actually be a face-on S0 galaxy (Capaccioli\markcite{m:c87} 1987;
Nieto\markcite{n:jl89} 1989) or at least a flattened elliptical (Strom
et al.\markcite{sesetal76} 1976; Statler\markcite{s:ts94} 1994).  Its
intrinsic shape was found to be almost oblate in the inner R/R$_e <$
0.4 (R$_e$ \app\ 54\arcsec) region, with axial ratios of b/a = 0.9 and
c/a = 0.5, and triaxial in the outer regions (b/a = 0.75 and c/a = 0.5)
(Capaccioli et al.\markcite{mcetal91} 1991, hereafter C91). These
authors also remarked the presence of a small stellar disk in the
central regions of the galaxy.  Combining photometric and kinematic
data with dynamical models, Statler\markcite{s:ts94} (1994) concluded
that this object is in fact an elliptical galaxy, ruling out very
flattened, triaxial shapes, but Statler \& Smecker-Hane\markcite{ssh99}
(1999), hereafter SS99, using more detailed stellar kinematic data,
suggested that the two-dimensional velocity field of \nttsn\ indicates
a two-component structure for the galaxy, with kinematical features
closely resembling those seen in the rotation curve of the edge-on S0
galaxy NGC\,3115.

Macchetto et al.\markcite{fdmetal96} (1996) (hereafter Paper I) mapped
the ionized gas of about 80 early-type galaxies, using
\halfa+[\ion{N}{2}] narrow-band imagery and found that \nttsn\ presents
a small (20\arcsec\ \vzs\ 14\arcsec) central disk of ionized gas with
an estimated mass of 2.2 \vzs\ \tento{3} \msol\ calculated from the
\halfa\ luminosity. Recent high spatial resolution photometric and
kinematic studies (Magorrian et al.\markcite{jmetal98} 1998; van der
Marel\markcite{vdm:rp99} 1999) showed that this galaxy could harbor a
central Massive Dark Object (MDO) of \app\ \tento{8}\ \msol. The best
fitting 3-integral models of Gebhardt et al.\markcite{kgetal96} (1996)
also indicates, besides the presence of a central MDO, that the
intrinsic shape of the galaxy would be an E6 seen at an inclination of
27\arcdeg.

In this work we address the stellar and gas kinematics, and the dust
morphology in the central two kpc of \nttsn\, using new spectroscopic
observations taken as part of our program to study the interstellar
medium in early type galaxies, as well as archival HST images. This
paper is organized as follows: Section~\ref{sect:obs} presents the
observations and data reduction. The luminosity profiles and dust
distribution derived from HST images are discussed in Section
\ref{sect:phot}. The stellar kinematics is explored in
Section~\ref{sect:stars}, where we test the simple case of disk
rotation in a spherical potential, and the gas kinematics is discussed
in Section~\ref{sect:gas}. The conclusions are summarized in
Section~\ref{sect:conc}. To allow direct comparison with our previous
results (Paper I; Ferrari et al.\markcite{ffetal99} 1999) we adopt
\hbc\ = 55 \kms Mpc$^{-1}$, and a distance to \nttsn\ of 14.5 Mpc,
although a number of different values can be found in the literature
(Sakai et al.\markcite{ssetal97} 1997; Ciardullo, Jacoby \&
Ford\markcite{cjf89} 1989). At this distance, 1\arcsec\ corresponds to
70 pc in the plane of the sky \footnote{For a distance of 10 -- 11 Mpc, the linear scale is 50 pc arcsec$^{-1}$, and the masses calculated in Section~\ref{sect:phot} would be overestimated by a factor of two.}.

\section{Observations and Data Reduction. \label{sect:obs}} 

\subsection{The HST images} \label{sect:imobs}

\begin{figure}
\psfig{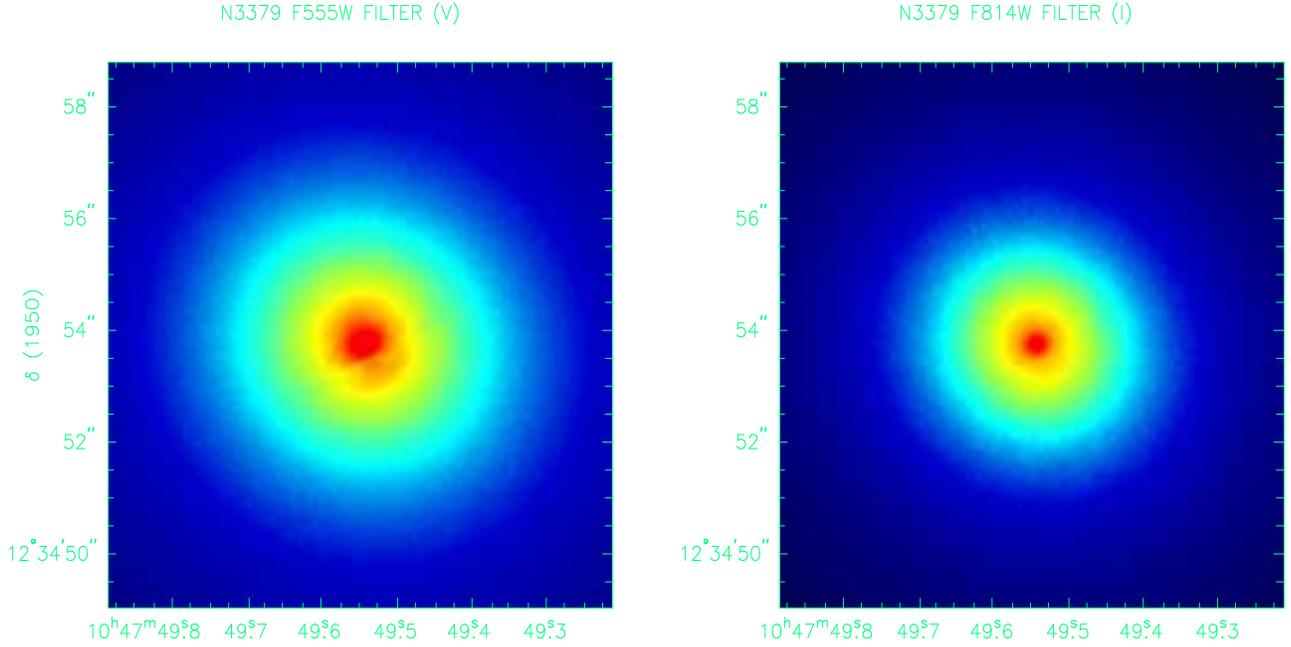}
\figcaption[n3379_f1.ps]{The HST/WFPC2 V and I images of NGC3379. Note the
dust lane cutting across the inner 2\arcsec, and the almost perfectly
circular appearance of the outer regions. North is up and East to the
left. \label{VIimages_fig}}
\end{figure}

Archival HST V and I images were used to analyze in detail the
photometric properties of the central ($r <$ 15\arcsec) region of
\nttsn. The images were obtained with the Wide Field Planetary Camera 2
(WFPC2) $+$ F555W and F814W filters, with the nucleus of the galaxy
placed on the PC frame resulting a spatial scale of 0.046 arcsec/pixel
and a resolution of \app\ 0\farcs1. The details of the observations
are given in Table~\ref{log_hst}. Data reduction of the individual
images followed the standard pipeline procedures for flat fielding and
bias correction. The individual images for each filter (four images with
the F555W and two with the F814W) were combined to remove cosmic rays,
resulting a total exposure time of 1660s and 800s for F555W and F814W
final images, respectively. Those were then flux calibrated in
order to convert the F555W  and F814W counts to the Johnson system at V
and I (Biretta et al.\markcite{jbetal96} 1996). The V and I HST
integrated magnitudes in a 10\arcsec\ aperture centered in the nucleus
agree within 0.13 and 0.15 mag, respectively, with the ground-based
data, implying very small uncertainties when deriving relative color
maps of the galaxy. The final F555W and F814W images are shown in
Figure~\ref{VIimages_fig}.

\subsection{Spectroscopy} \label{sect:spobs}

\begin{figure}
\plotone{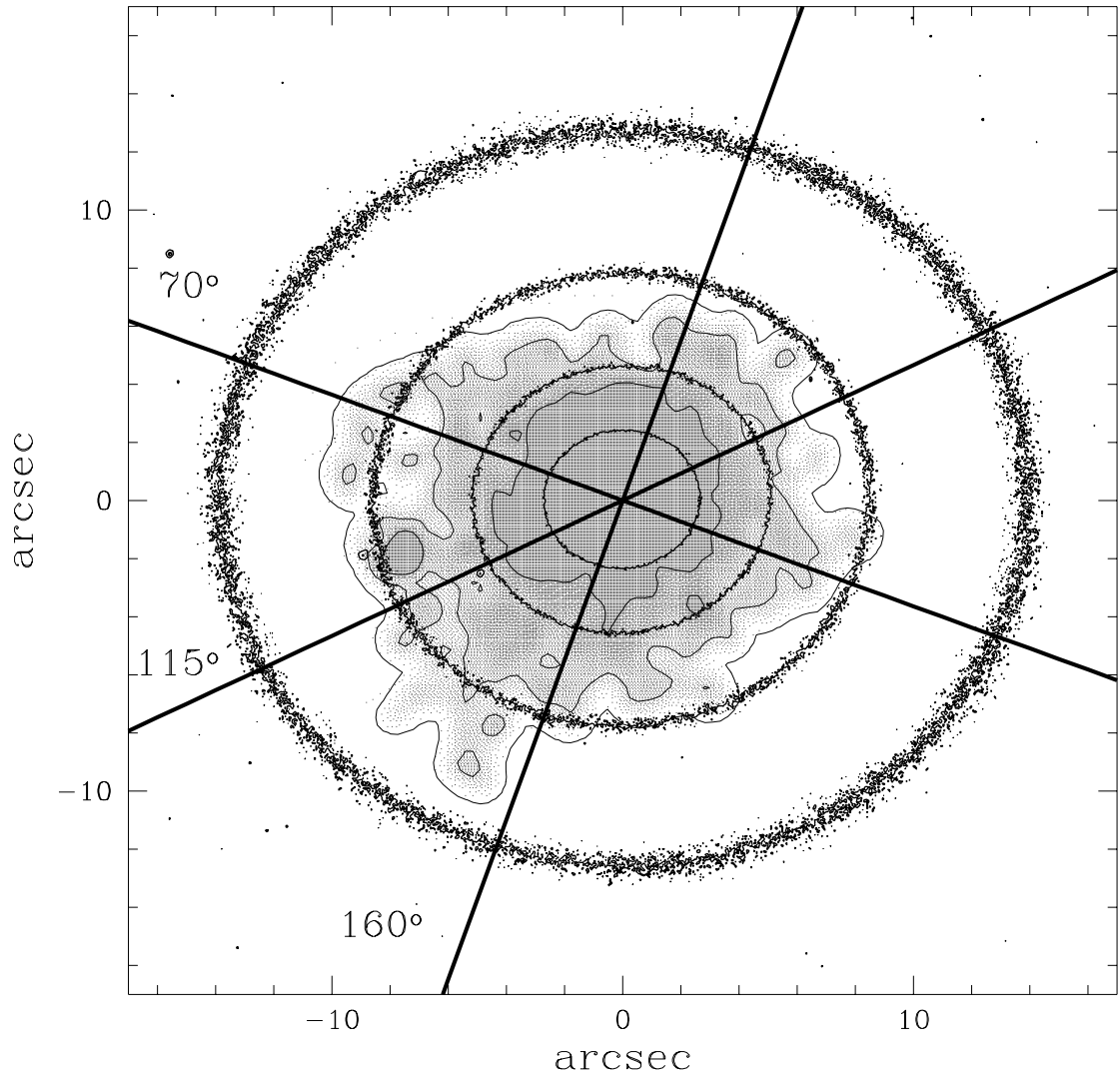}
\figcaption[n3379_f2.eps]{The observed slit positions superimposed on the
\halfa$+$[\ion{N}{2}] image from Macchetto et al. (1996) and V-band
isophotes map from the HST image. Orientation is the same as in
Figure~\ref{VIimages_fig}. \label{slits_fig} }
\end{figure}

To study the  kinematics of the ionized gas detected in Paper I and to
explore earlier suggestions about the presence of a stellar disk in the
inner 30\arcsec\ of \nttsn\ (C91; SS99), long-slit spectra were
obtained at position angles 70\arcdeg,  115\arcdeg, and 160\arcdeg,
corresponding respectively to the major, intermediate,  and minor axes
of the gas distribution. The slit positions are shown in
Figure~\ref{slits_fig}, superimposed on the \halfa$+$[\ion{N}{2}] image
and R-band isophotal map of \nttsn\ from Paper I.

The observations were conducted at the ESO 3.6m telescope $+$ EFOSC1 at
La Silla, Chile. In order to maximize the coverage of the relevant
stellar absorption and gaseous emission lines, we used the O150 grism
in the range 5140--6900 \AA\ with a dispersion of 3.4 \AA/pixel and the
CCD \#26 (Tektronik 512 \vzs\ 512 pixels), with a spatial scale of 0.61
arcsec/pixel and \app\ 1\arcsec\ resolution.  The slit width was set to
1\arcsec, with an useful length of 3\arcmin.  The observational
procedure was to obtain two exposures of 40--45 minutes at each
position angle in order to increase the S/N ratio and facilitate the
removal of cosmic rays. The observations are summarized in
Table~\ref{log_eso}. The stability of the instrument has been checked
by comparing the position of the sky emission lines present in the
galaxy spectra, and we found that the relative shift is always less
than 0.2 pixels, with a rms of about 0.1 pixels (\app\ 17 \kms). We
also verified the correct alignment of the CCD Y axis with the
dispersion direction by tracing, for each spectrum, the position of the
centroid of the galaxy or star light distribution along each detector
row. No systematic misalignment was found, with the tilts never larger
than 0\farcs05 (0.5 pixels).  Several kinematic (giant stars in the
spectral type range K0-K4 III, with rotation velocity less than 20
\kms) and spectrophotometric standards were observed at the beginning
and the end of each night.  The data reduction was carried out the
usual way, using IRAF\,\footnote{IRAF is distributed by the National
Optical Astronomy Observatories, which are operated by the Association
of Universities for Research in Astronomy, Inc., under cooperative
agreement with the National Science Foundation.} tasks.  The mean bias
level was subtracted and the frames divided by a normalized dome
flat-field to remove pixel-to-pixel variations. The averaged sky flat
was used to correct for the non-uniformity of the response along the
slit.  Wavelength calibration was obtained fitting a low order
polynomial to the centroid of the brightest, non-saturated lines in the
HeAr comparison lamp spectra, with residuals of less than 0.15 \AA.
Relative velocities were measured with an uncertainty of \app\ 5 \kms.

A cross correlation algorithm, working in pixel space, has been
used to derive the stellar kinematics (Dalle Ore et
al.\markcite{cdoetal91} 1991). Briefly, the continuum of the galaxy and
template star spectra were fitted with a polynomial of order \app\ 20,
and subtracted.  Each resulting galaxy spectrum was then
cross-correlated with that of the kinematic template star, and the
center of a Gaussian fitted to the central peak of the correlation
function gives the radial velocity of the galaxy with respect to the
template star.

\section{Results}

\subsection{Luminosity profiles and \vmi\ image} 
\label{sect:phot}

\begin{figure}
\plotone{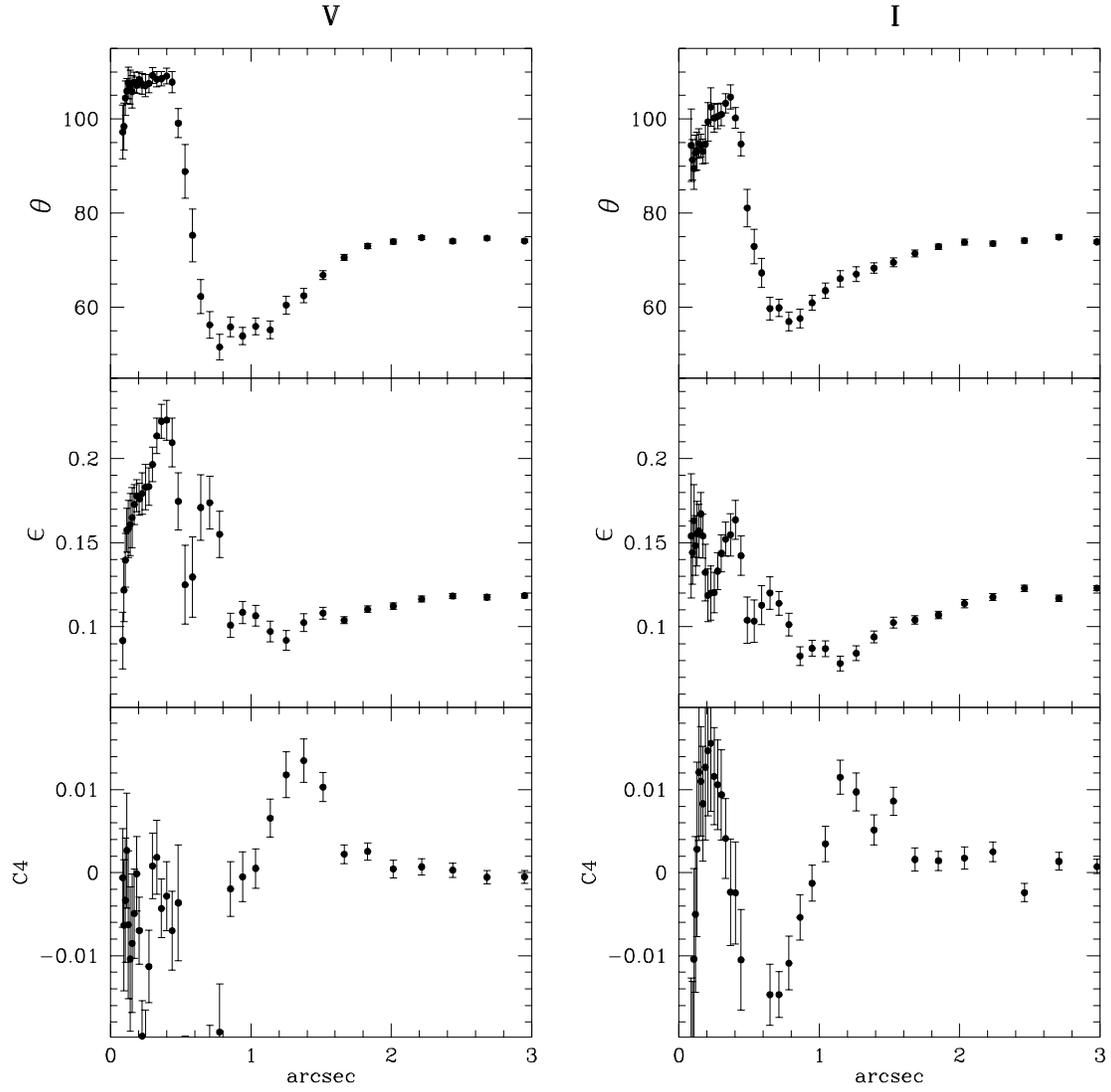}
\figcaption[n3379_f3.eps]{Results of the isophote fitting as a function of the
distance along the major-axis {\it a} in arcsec. From top to bottom:
position angle of the major axis, ellipticity, and Fourier
\fcf\ parameter. \label{isofit_fig}  }
\end{figure}

Elliptical galaxies have central surface-brightness
distributions of two distinct types: those with steep profiles that do
not show a break in the slope -- called ``power-law'' galaxies; and
those that do show a break at a given radius $r_b$, from a steeper
outer profile to a flatter core -- called ``core'' galaxies (Lauer et
al.\markcite{tletal92} 1992, \markcite{tletal95} 1995; Crane et
al.\markcite{pcetal93} 1993; Ferrarese et al.\markcite{lfetal94} 1994;
Forbes, Franx \& Illingworth\markcite{ffi95} 1995; Faber et
al.\markcite{sfetal97} 1997). \nttsn\ was classified by Faber et
al.\markcite{sfetal97} (1997) as a ``core'' type galaxy, and in
addition to the characteristic luminosity profiles, we found that the
HST V and I images reveal a well defined nuclear dust structure (see
Figure~\ref{VIimages_fig}), similar to that observed in the E2 galaxy
NGC\,6251 (Ferrarese \& Ford\markcite{ff99} 1999). The  presence of
such nuclear dust disk raises the possibility that the observed central
luminosity profile could be altered by absorption, creating an
artificial break and mimicking a ``core'' type profile. To analyze the
role that the dust plays on the luminosity profile, we carried out
detailed surface photometry of the central region of \nttsn\, using the
IRAF task {\tt ellipse} to measure the main isophotal parameters
(position angle, ellipticity and Fourier coefficient \fcf). The fitting
results are shown in Figure~\ref{isofit_fig} as a function of the
distance $a$ in arcsec, along the isophotal major axis.

The position angle of the major axis presents an almost constant value
of 75\arcdeg\ between 2\arcsec\ and 13\arcsec, decreasing slightly to
70\arcdeg\ for larger distances.  Within the inner 2\arcsec, on the
other hand,  the PA changes considerably, first decreasing to lower
values of up to 50\arcdeg\ at $a$ \app\ 1\arcsec, and then increasing
again to reach a value of PA = 120\arcdeg\ at 0\farcs4, which is
close to the orientation of the dust disk's major axis.  The isophotes
external to 5\arcsec\ show a constant ellipticity with a value of
0.12, but internal to this radius the ellipticity increases to up to
0.24 at 0\farcs5.  It is very significant that the changes in both
the PA of the major axis and the ellipticity of the isophotes start at
approximately 2\arcsec,  in close correspondence with the presence of
the dust disk.

The \fcf\ parameter is related to the shape of the isophotes:
ellipticals with ``core'' profiles tend to present ``boxy'' isophotes
(\fcf\ $<$ 0), while ``power-law'' galaxies tend to be more disk-like
(\fcf\ $>$ 0) (Nieto, Bender \& Surma\markcite{nbs91} 1991). In \nttsn,
the \fcf\ parameter external to 2\arcsec assumes an almost constant
value of zero, while internal of this radius it first rises to positive
values up to $a$ \app\ 1\farcs2 and then oscillates around zero in the
innermost regions. Such behaviour indicates that, outside the region
where the dust disk is present, this galaxy has indeed almost perfectly
elliptical isophotes, and that the dust affects the very inner
isophotes making them appear more disk-like where they would be
expected to be ``boxy'' based on the classification of \nttsn\ as a
``core'' galaxy (Faber et al.\markcite{sfetal97} 1997;
Kormendy\markcite{k:j99} 1999).

The  derived V and I luminosity profiles along the major axis at PA =
70\arcdeg\ show evidence of a  change in slope at $r$ \app\ 0\farcs7.
In order to verify if this depression could be attributed to the
presence of the dust disk, we performed the {\tt ellipse} fitting with
the region occupied by the disk masked-out, and found the same profiles
as in the previous case  (without masking). Therefore, the presence of
dust has a negligible effect on the profiles.  The change in the
profile slope makes necessary to use two different laws for the fitting:

-- from 0\farcs08  to 0\farcs68, where the profiles are shallower, a
cuspy law was used:

\begin{eqnarray}
 I & = & I_0 \, r^{-\gamma}  \label{eq:core} \\ 
 \mu-\mu_0 & = &-\gamma\log(r)  \nonumber  
\end{eqnarray}

-- from 0\farcs68 to 15\arcsec, the S\'ersic law
(S\'ersic\markcite{s:j68} 1968) provides a better fit:

\begin{eqnarray}
I(r) &=& I_{n} \; \exp \left\{ -b_{n} \left[ \left( \frac{r}{r_{n}}\right)
 ^\frac{1}{n} -1 \right] \right\}  \label{eq:sersic} \\
b_{n} &=& 0.868 n - 0.142 \nonumber
\end{eqnarray}

The resulting parameters of the fit are shown in Table \ref{phot_tab}. 

\begin{figure}
\plotone{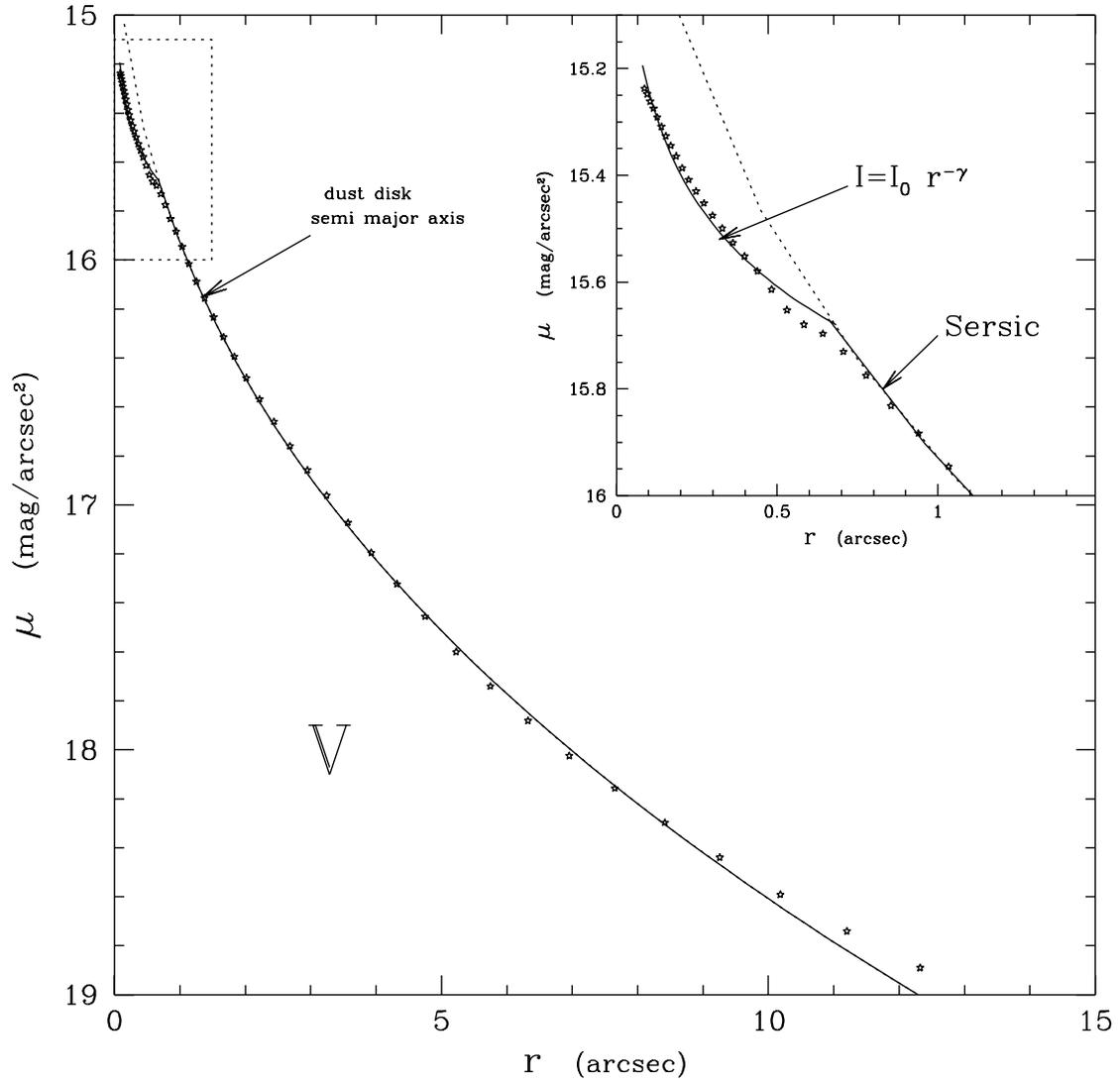}
\figcaption[n3379_f4.eps]{Luminosity profile decomposition for the HST V
image. The insert to the right shows an expanded view of the central
1\farcs5, where the ``cuspy'' law dominates. \label{photV_fig} }
\end{figure}

Figure~\ref{photV_fig} shows the observed (star symbols) and fitted
profiles (solid line) from the V image. The interval 0\farcs7
$\lesssim r \lesssim$ 9\arcsec\ is well reproduced by a S\'ersic law
with  index $n=2.36$, indicating that the profile is more concentrated
than the usual de Vaucouleurs light distribution, which fails to
reproduce the observed luminosity.  The very inner region ($r<$
0\farcs7) is expanded in the zoomed insert to the right, where it
can be clearly seen where the S\'ersic law breaks up and the nuclear
profile dominates.

\begin{figure}
\psfig{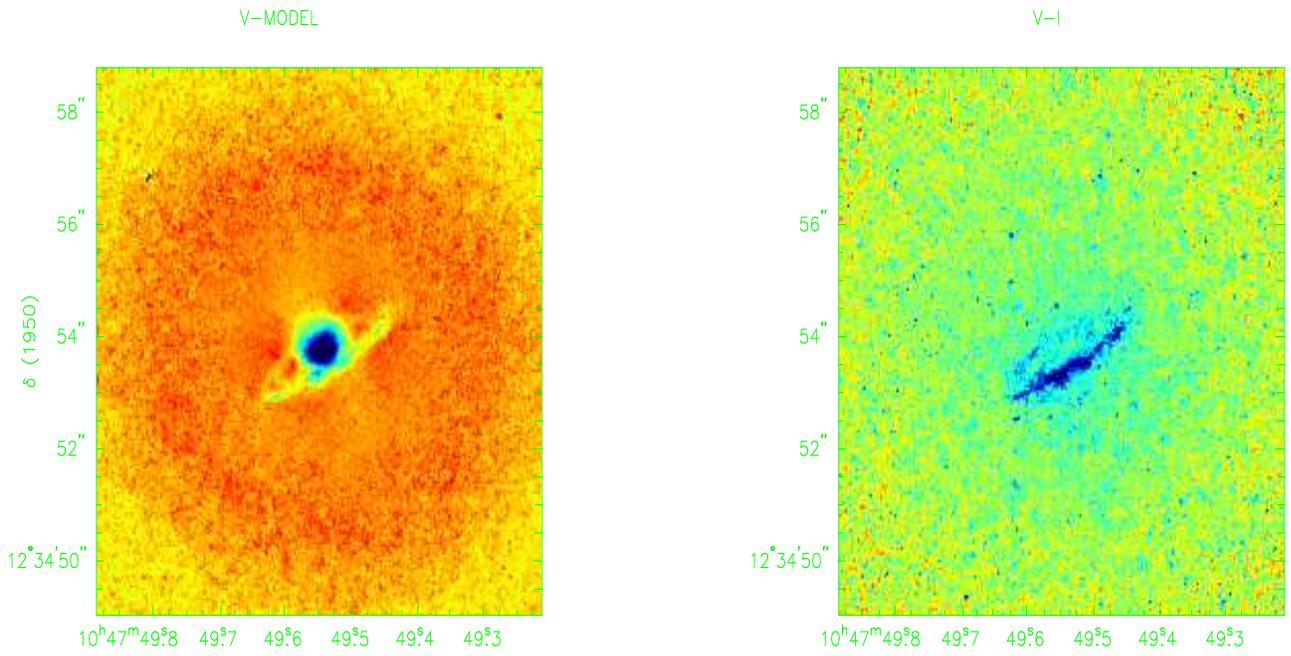}
\figcaption[n3379_f5.ps]{Left: the HST V image after the subtraction of the
bulge model as described in the text. Note the central cusp, where
the luminosity profile is fainter than the S\'ersic law, and the dust
ring with $r$ \app\ 1\farcs5.  Right: \vmi\ color-index image. Only the
dust ring is visible, evidencing that the central cusp is not an effect
of localized absorption. Orientation as in Figure \ref{slits_fig}.
\label{colourVI_fig}  }
\end{figure}

The fitted S\'ersic profile was assumed to represent the stellar light
distribution for $r \lesssim$\ 10\arcsec, and a model image was built
using the IRAF task {\tt bmodel}.  This model was then subtracted from
the V image and the result is shown in the left panel of
Figure~\ref{colourVI_fig}.  For $r\lesssim $ 0\farcs7, where the
``cuspy'' law dominates, the nuclear core appears now very clearly in
absorption, since its profile is  $\simeq$\ 0.2 mag fainter than the
external profile extrapolation.  The subtraction also reveals a very
conspicuous dust structure at $r\simeq$\ 1\farcs5, reminiscent of the
dust disk seen in NGC\,6251 (Ferrarese \& Ford\markcite{ff99} 1999).
The \vmi\ color-index image, shown in the right panel of
Figure~\ref{colourVI_fig}, confirms the presence of dust disk with
major and minor axes 3\arcsec and 0\farcs8 long, respectively, with the
major axis oriented along PA \dapp\ 125\arcdeg. Assuming that the disk
is circular, the observed axial ratio implies an inclination of
77\arcdeg\ between the line of sight and the normal to the disk.

From our data, we measured a mean $A_{\rm V}$ extinction of 0.11 mag
inside the disk. Assuming that the dust is in a uniform layer with
$r\simeq$ 1\farcs5 (105 pc) and is composed mainly of silicate grains
with mean size 0.08\,$\mu$m, a dust mass of 150 \msol\ was estimated
following the method discussed in Ferrari et al.\markcite{ffetal99}
(1999). The analysis in that paper also indicated presence of a more
extended, asymmetric dust cloud extending about 5\arcsec\ to the South,
with an estimated mass of \app\ 100\msol, and mean $A_{\rm V}$
extinction of 0.027 mag. The much larger mass and extiction measured in
the disk over that of the southern extended dust cloud, indicates the
presence of a dust gradient towards the central regions of the galaxy.

\begin{figure}
\plotone{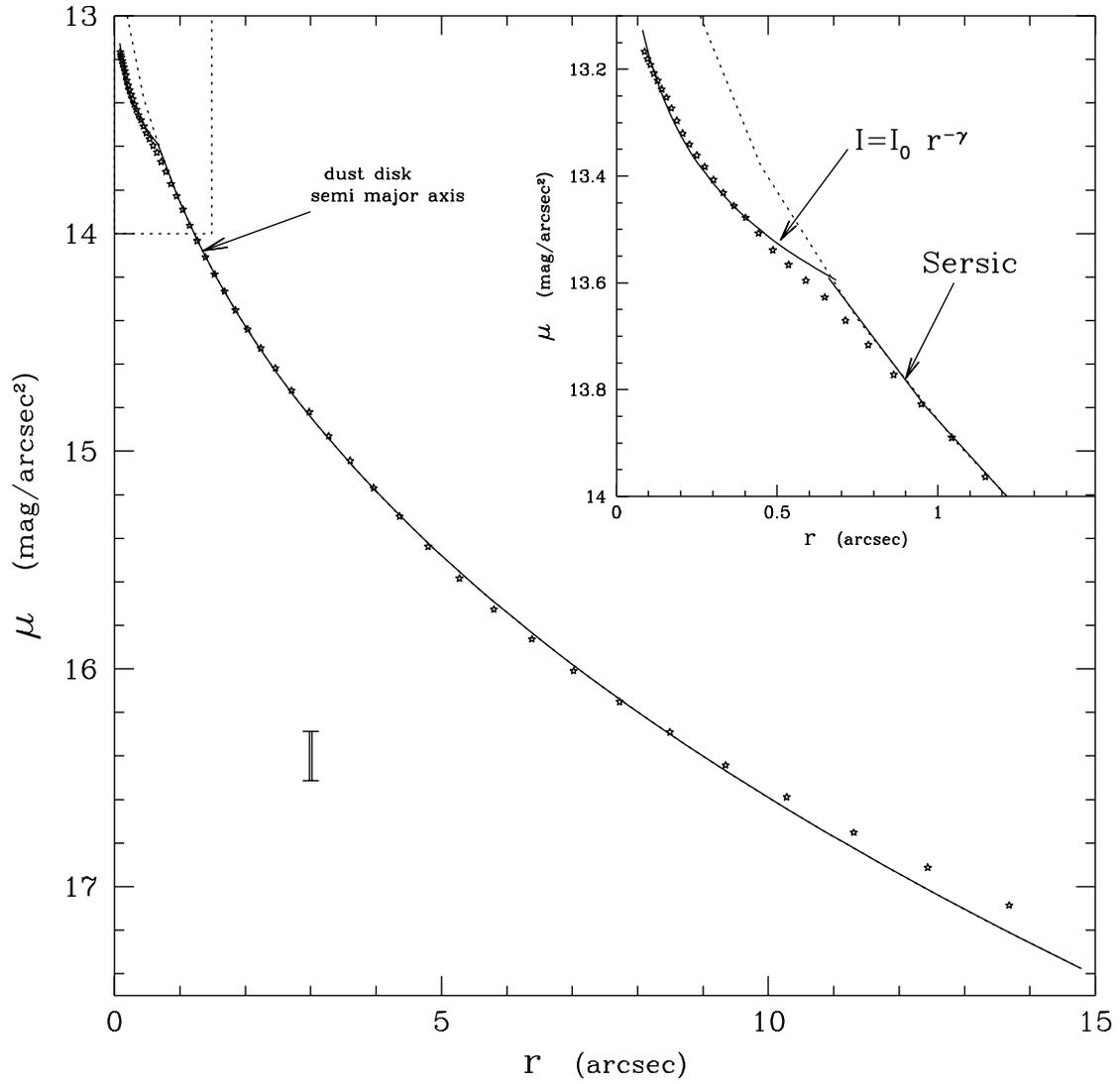}
\figcaption[n3379_f6.eps]{Luminosity profile decomposition for the HST I
image.  \label{photI_fig} }
\end{figure}

The fit of the I band luminosity profile yields very similar results,
as can be seen in Table \ref{phot_tab} and Figure \ref{photI_fig}.
Since the I band should be less affected by reddening, the presence of
the central cusp in the I profile shows that absorption by dust cannot
be the primary cause for the observed change of slope.  The isophotal
parameters discussed above indicate that dust indeed plays a role in
distorting the inner isophotes, but it is not a strong enough effect to
modify the global luminosity profiles.

This galaxy therefore presents a characteristic ``cuspy'' core, with
the steeper outer (S\'ersic) profile breaking up at $r_b$ \dapp\ 0\farcs7
to be replaced by a shallower inner profile, given by Eq.~\ref{eq:core}
with $\gamma$ = 0.21. Note that the profile break is located
about 0\farcs8 from the most absorbed part of the dust disk, and
therefore the light depletion observed in the luminosity profile cannot
be explained by dust absorption, since it is not present where this
latter is stronger. This same conclusion is reinforced by the
\vmi\ color-index image (Fig.~\ref{colourVI_fig}), where only the dust
ring is visible, evidencing that the central cusp is not an effect of
localized absorption.

Several models have been proposed to account for core formation, such
as: a) an isothermal sphere with density profile $r^{-2}$; b) black
holes (BH) adiabatically grown in homogenous isothermal cores
(Young\markcite{y:p80} 1980; Van der Marel\markcite{vdm:rp99} 1999); c)
the orbital decay of massive black holes acreted in mergers, where the
decaying BHs may heat and eject stars from the center, eroding the
power-law if any exists and creating a core (Faber et
al.\markcite{sfetal97} 1997 and references therein); or d) remnants
formed by mergers of disk galaxies, where the dissipation in the gas
and ensuing star formation during the merger process leaves a dense
stellar core in the remnant (Mihos \& Hernquist\markcite{mh94} 1994).
These last models can account for the extra light at small radius (over
a S\'ersic or de Vaucouleurs functions) in the luminosity profile of
``power-law'' type galaxies (Kormendy\markcite{k:j99} 1999),  but
cannot  reproduce the broken light profile observed in \nttsn, which is
fainter than the S\'ersic distribution in the inner region.

Magorrian et al.\markcite{jmetal98} (1998) constructed dynamical models
for a sample of 36 early-type galaxies, including \nttsn, using HST
photometry and ground-based kinematic data. The models assume that the
galaxies are axisymmetric, described by a two integral distribution
function, with an arbitrary inclination angle, a position-independent
stellar mass-to-light ratio, and a central massive dark object (MDO) of
arbitrary mass. Such approach was able to provide acceptable fits to 32
galaxies in the sample, and a model with an MDO mass of 3.9
\vzs\ \tento{8} \msol\  and a mass-to-light ratio $\Upsilon$ of 5.3 was
found to be an adequate description for the stellar kinematics of
\nttsn. In a similar approach, Gebhardt et al.\markcite{kgetal96}
(1996) used axisymmetric, 3-integral models to fit both HST/FOS and
ground-based spectroscopy along the major and minor axes. HST/WFPC2 and
ground-base images were used to constrain the light distribution. The
best fit model for \nttsn\ also required the presence of a central MDO
with mass of 6 \vzs\ \tento{8}\ \msol.

Recently, van der Marel\markcite{vdm:r99} (1999) studying the models
first proposed by Young\markcite{y:p80} (1980) of adiabatically grown
black holes, predicted that the luminosity profiles should behave as $I
=  r^{-1/2}$ for asymptotically small radii, but as $I = r^{-\gamma}$
at radii observable with HST, with the index $\gamma$ increasing
monotonically with the mass of the central object. The index $\gamma$
can assume all the observed values and therefore, both ``core'' and
``power-law'' profiles can be reproduced. Within this approach, the
luminosity profile  of \nttsn\ can be reproduced with the inclusion of
a central black hole of mass of 1.9 \vzs\ \tento{8}\ \msol.

\subsection{Stellar Kinematics: an inner disk?}
\label{sect:stars}

\begin{figure}
\plotone{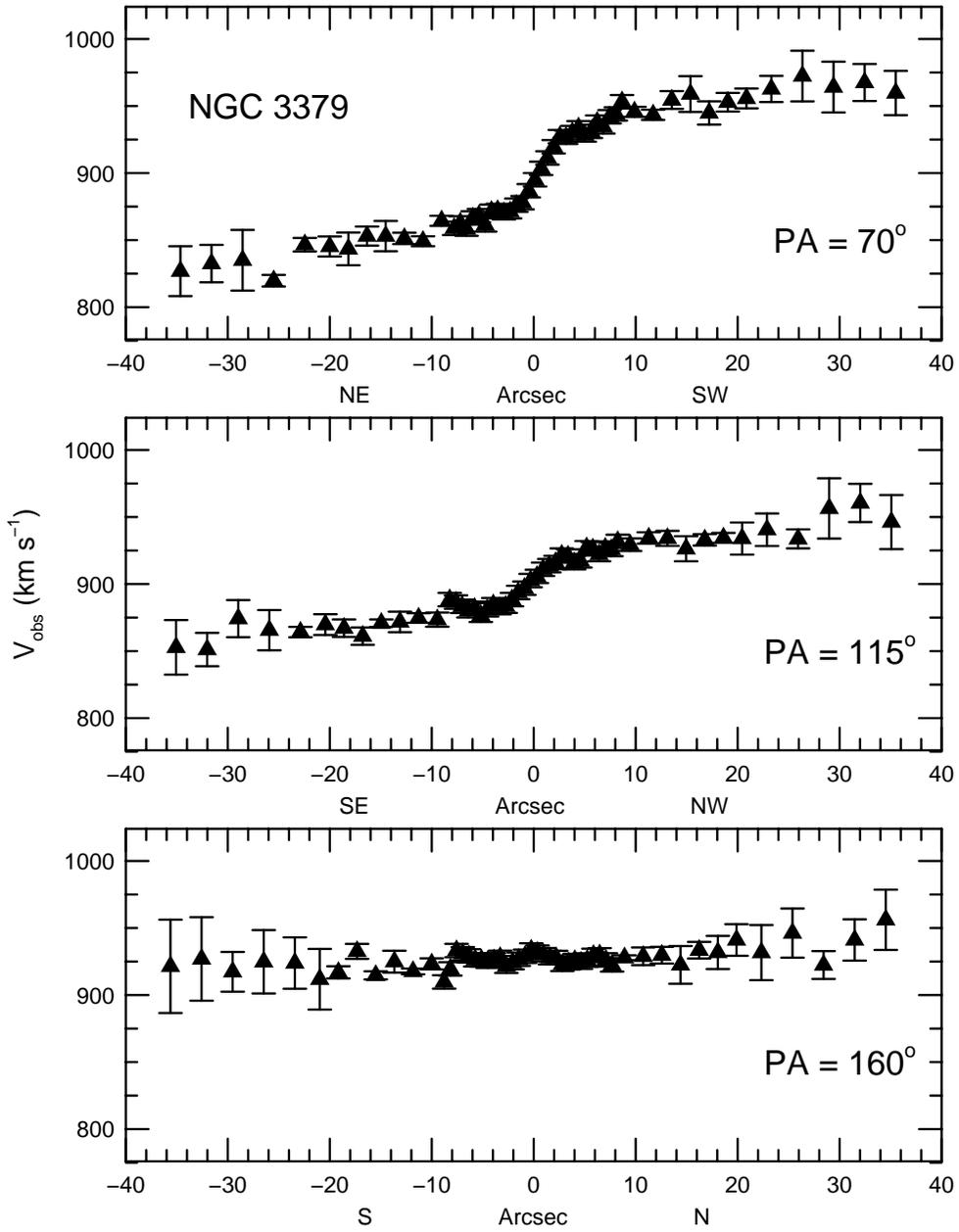}
\figcaption[n3379_f7.ps]{The observed rotation curves for PA = 70\arcdeg, 
115\arcdeg, and 160\arcdeg. \label{starrot_fig} }
\end{figure}

In this section, we use the spectroscopic data described in Section
\ref{sect:spobs} to explore the possible presence of a stellar disk as
well as to study the kinematics of the ionized gas shown in Paper I.
The observed stellar rotation curves are presented in
Figure~\ref{starrot_fig}.  In agreement with previous results (Franx at
al.\markcite{fih89} 1989; Bender et al.\markcite{bsg94} 1994; SS99), it
can be seen that the galaxy shows a fairly symmetric rotation curve
along PA = 70\arcdeg and 115\arcdeg, with a steeper velocity gradient
observed in the inner 6\farcs The rotation curve along PA =
160\arcdeg\ is almost flat, confirming once more the presence of
rotation around the galaxy's minor axis. The kinematic breaks remarked
by SS99 at \app\ 5\arcsec\ are also clearly discernible in our data.

To model the stellar kinematics, we assumed that the stars follow
circular orbits in a plane, and used  Bertola et al.\markcite{fbetal91}
(1991) analytic expression as a first order approximation for the
observed rotation curve:

\begin{equation}
V_c(r) = \frac{AR}{(R^2 + C_o^2)^{p/2}} + V_{sys}
\label{eq:rot}
\end{equation}

where $V_{sys}$ is the systemic velocity, $R$ is the radius in the
plane of the disk and $A$, $C_o$, and $p$ are parameters that define
the amplitude and shape of the curve. If $v(r,\Psi)$ is the observed
radial velocity at a position $(r,\Psi)$ in the plane of the sky, where
$r$ is the projected radial distance from the nucleus and $\Psi$ its
corresponding position angle, we have:

\begin{equation}
v_{mod}(r,\Psi) = V_{sys} + \frac{A\,r\,\cos(\Psi - 
			\Psi_o)\,\sin i\,\cos^p i}
{\{r^2\,\eta + C_o^2\,\cos^2 i \}^{p/2}}
\end{equation}

\noindent 
with
\begin{displaymath}
\eta \equiv [\sin^2 (\Psi - \Psi_o) + \cos^2 i\,\cos^2(\Psi - \Psi_o)]
\end{displaymath}

where $i$ is the inclination of the disk ($i=0$ for a face-on disk) 
and $\Psi_o$ the position angle of the line of nodes.

\begin{figure}
\plotone{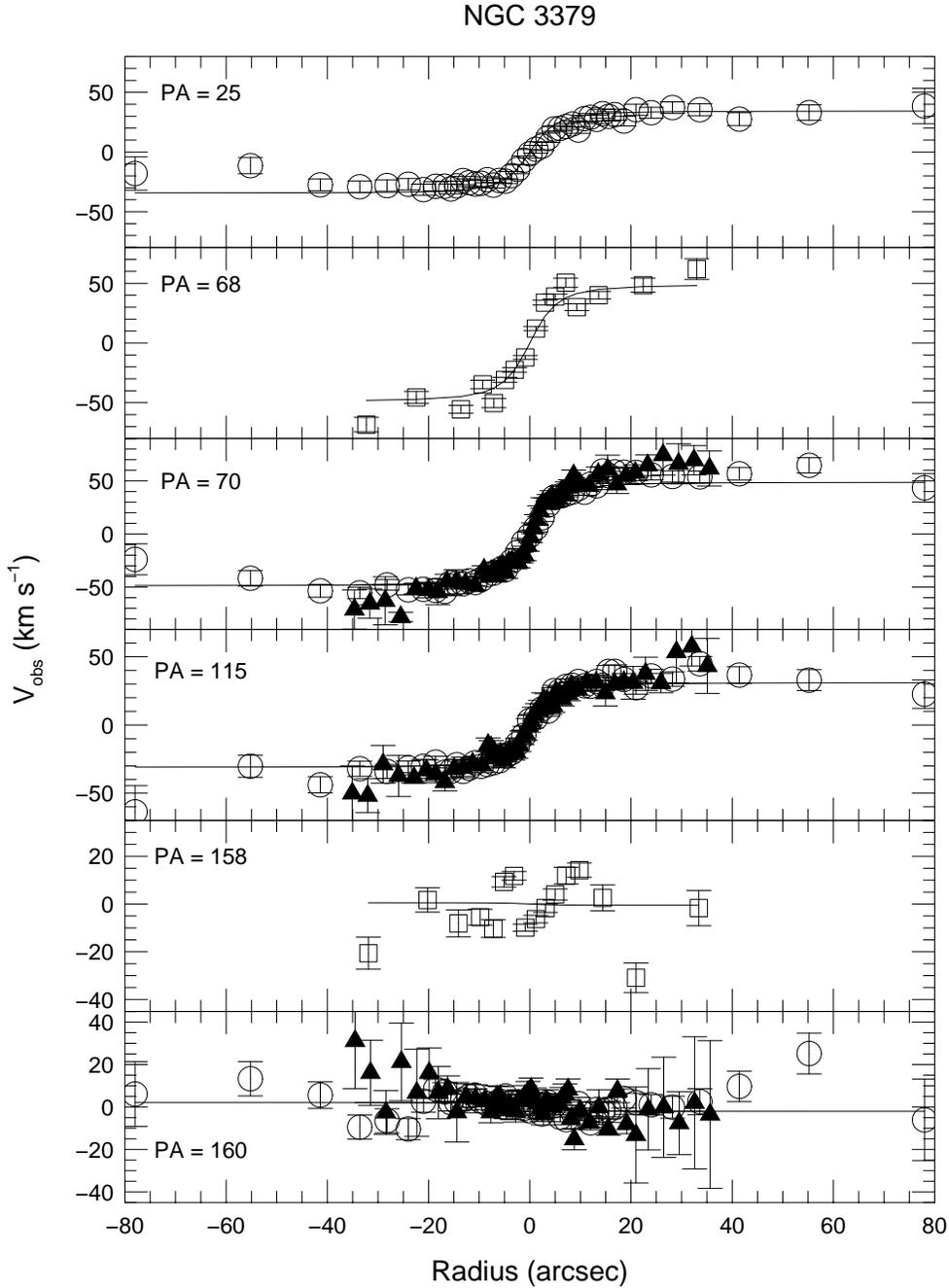}
\figcaption[n3379_f8.ps]{The observed stellar rotation curves and the
resulting planar disk model fit (solid line).  The filled triangles
represent the data presented in this paper, open squares are the data
from the two positions angles published by Franx et al. (1989), and the
open circles represent the data from SS99. \label{starfit_fig} }
\end{figure}

The stellar rotation curves for PA = 70\arcdeg, 115\arcdeg, and
160\arcdeg\ described above were combined with the data from Franx et
al.\markcite{fih89} (1989) for PA = 68\arcdeg\ and 158\arcdeg, and from
SS99 for PA = 25\arcdeg, 70\arcdeg, 115\arcdeg, and 340\arcdeg.  Our
data agree very well with the rotation curves given by SS99 along the
same PAs (see Figure \ref{starfit_fig}). The various parameters were
determined simultaneously by minimizing the residuals $\Delta v =
v_{obs} - v_{mod}$, with $v_{mod}(r,\Psi,A,C_o,p,i)$ and
$v_{obs}(r,\Psi)$ being the model and observed radial velocities at the
position $(r,\Psi)$ in the plane of the sky, respectively.

We used a Levenberg-Marquardt non-linear least-squares algorithm to fit
the above model. The data from the nine sets were first fitted with $A,
C_o, p, \Psi_o, i$ and $V_{sys}$ as free parameters, with $V_{sys}$
having the same value for all data points (except for the SS99 data
where $V_{sys}$ was inittially set to zero). Then, to take into account
possible zero point velocity offsets between the data sets, each one
was separately fitted allowing $V_{sys}$ to vary, but keeping the
remaining parameters fixed to the values obtained before. The resulting
systemic velocity for each set was subtracted from the observed values
and all data refitted with $A, C_o, p, \Psi_o$ and $i$ as free
parameters. Finally, the effect of the initial guess on the parameters
was explored running a Monte Carlo search for a wide interval of values
for each parameter, while the others were given their best fit values
as the initial guess. An unstable fit will show as a large deviation
from the original model.

The best solution (Table \ref{starrot_tab} and solid line in Fig.
\ref{starfit_fig})  corresponds to an asymptotically flat rotation
curve.  The model is  a good representation for the data up to R
\app\ 20\arcsec, and indicates that the stellar  disk is inclined
\dapp\ 26\arcdeg\ relative to plane of the sky, with the apparent major
axis located at PA \dapp\ 67\arcdeg. These values are  in very good
agreement with  the V and I isophotal parameters analyzed in
Section~\ref{sect:phot}, where it was shown that the PA of the major
axis has almost constant value of 75\arcdeg\ between 2\arcsec\ and
13\arcsec, converging to 70\arcdeg\ for larger distances, and the
ellipticity of the isophotes external to 5\arcsec\ has a constant value
of 0.119, which (assuming that the isophotal contours originate in a
circular stellar disk) imply an inclination of \dapp\ 28\arcdeg.  Our
results are also in agreement with the parameters derived
photometrically by CP91. The observed deviation of the data points from
the planar rotation model at larger radii is also expected within C91's
analysis, where the triaxiality of the bulge grows outward and our
model of purely planar rotation cannot represent the more complex
stellar velocity field. We stress that our analysis does not {\it
prove} the existence of a stellar disk in \nttsn. It shows that such a
disk is indeed a plausible first order representation, but with the
limited spatial resolution of ground-based observations it is not
possible to fully deconvolve the triaxial motions of an elliptical or
bulge system and show that a planar disk is the {\it only} possible
model for the observed stellar kinematics.

\subsection{Gas kinematics}
\label{sect:gas}

\begin{figure}
\plotone{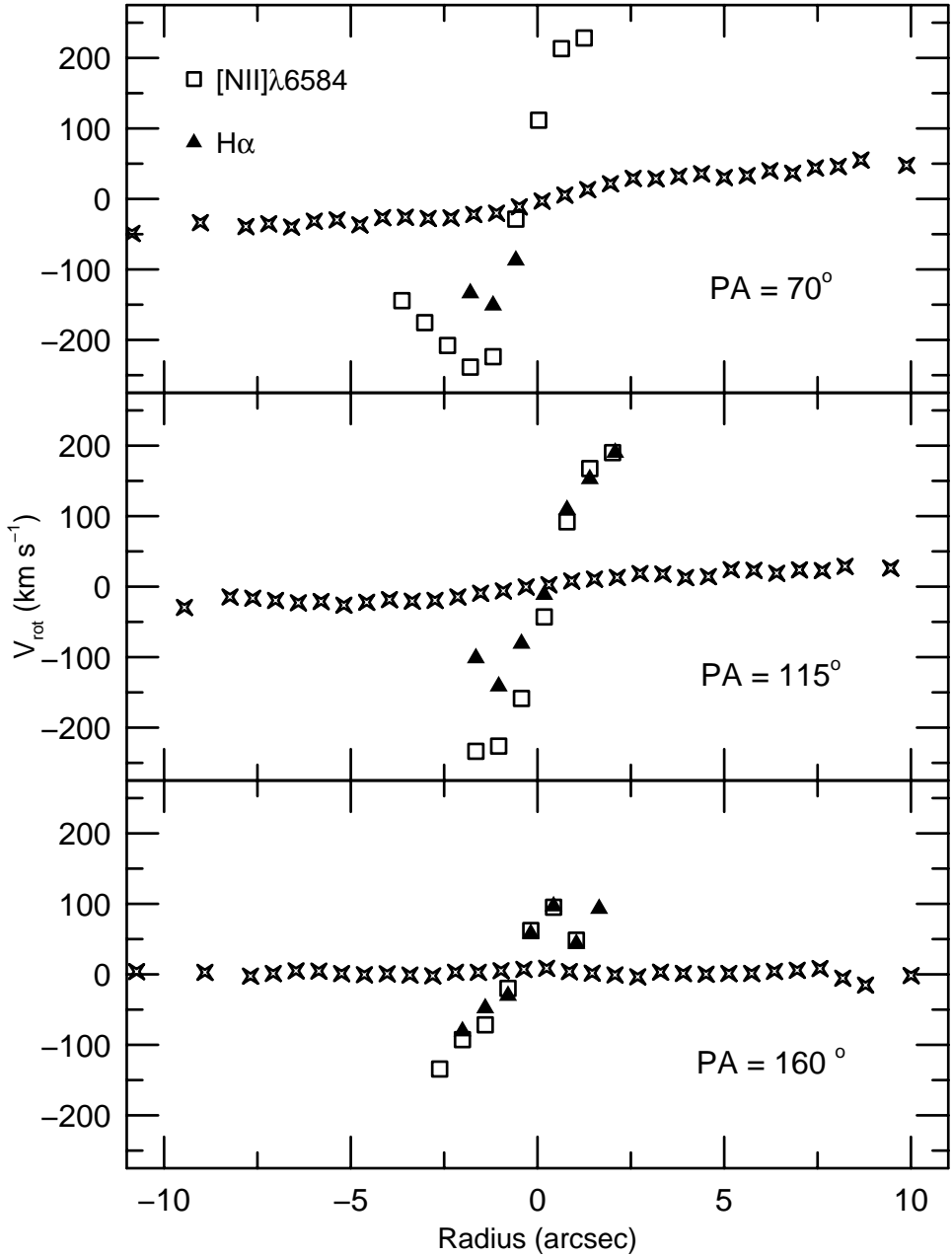}
\figcaption[n3379_f9.ps]{The observed emission line rotation curves
([\ion{N}{2}], open squares; \halfa, filled triangles), superimposed on
the correspondent stellar rotation curves (star symbols). Note the
much higher velocity of the gas. \label{gasrot_fig} }
\end{figure}

To analyze the gas kinematics, we have used the \halfa$+$\niit 6584
emission lines which are the only ones detected in our data. The
underlying stellar continuum was obtained integrating the spectra
outside the line-emitting region, and then subtracted from the central
rows. The gas velocity was measured by interactively fitting a single
Gaussian component to the emission lines in each individual spectrum.
Reliable measurements were obtained only in the inner 5\arcsec,
corresponding to approximately half the extension of the ionized gas
region detected in Paper I. The gas velocities along the three observed
positions are shown in Figure~\ref{gasrot_fig} superimposed on the
stellar rotation curves.

In the inner 5\arcsec, the observed velocities are of up to \app\ 240
\kms\ at PA=70\arcdeg , and up to \app\ 210 \kms\ at PA = 115\arcdeg; a
smaller rotation ($\Delta$v \app\ 90 \kms) is still observed at PA =
160\arcdeg. The steeper gradient of the gas rotation curves indicates
that it rotates much faster but in the same direction as the stars.
Since similar maximum rotational velocities are observed at both PA =
70\arcdeg\ and 115\arcdeg, we can quite safely assume that the position
angle of the major axis of the gas disk would be approximately halfway,
namely at PA \app\ 85\arcdeg. Assuming that the gas is indeed in a
disk, the axial ratio derived from the isophotal fitting in the
\halfa$+$[\ion{N}{2}] image implies a inclination $i$ of 25\arcdeg.
Using these two values for the projection angles, the deprojected
velocity of the gas at $r$ = 1\farcs3 (91 pc) is \app\ 580 \kms. If the
gas was in pure Keplerian rotation, this value would imply that the
total mass inside this radius is \app\ 7\vzs\ \tento{9} \msol.
Therefore, a quite large gravitational mass would be required inside
the inner 2\arcsec\ in order to support the gas rotation. The limited
amount of data, however, precludes a more detailed analysis.

\section{Conclusions}
\label{sect:conc}

From our analysis of archival V and I HST images and ground based 
spectra of \nttsn,  we conclude that:

\begin{enumerate} 

\item The V and I luminosity profiles in the inner 13\arcsec\ cannot be
represented by a single component, with the fitting revealing two
different systems: the external one following a S\'ersic Law with a $n$ =
2.36 index, and therefore steeper than the usual de Vaucouleurs
$r^{-1/4}$ profile; and a central core ($r < $ 0\farcs7) with a
characteristic ``cuspy'' profile.

\item The \vmi\ color index image and the subtraction of the underlying
stellar component represented by the fitted S\'ersic profile from the V
image revealed the presence of a small ($r$ \app\ 105 pc) dust disk,
oriented at PA \app\ 125\arcdeg\ and inclined $i$ \app\ 77\arcdeg\ with
respect to the line-of-sight, with mass \app\ 150\msol.

\item  The stellar rotation in the inner 20\arcsec\ can be quite well
described by a parametric disk model. The model indicates that the disk
is inclined \app\ 26\arcdeg\ relative to the sky plane, with the
apparent major axis located at PA \app\ 67\arcdeg. While our analysis
does not  unambiguously demonstrate the existence of a stellar disk in
NGC3379, it shows that within the limitations of the data,  rotation in
a planar disk is a plausible fit to the observed stellar rotation
curves.

\item The gas in the inner 5\arcsec\ rotates in the same direction, although much faster, than the stars.

\end{enumerate}

\acknowledgments

We acknowledge useful discussion with R. van der Marel and helpful
remarks from the anonymous referee. This work was supported in part by
the Visitor Program of  STScI, the brazilian Institutions CAPES and
CNPq, and grant FINEP/PRONEX 7697100300.






\newpage

\begin{deluxetable}{lcc}
\tablecolumns{3}
\tablewidth{164 pt}
\tablecaption{Journal of Observations -- HST/WFPC2 images 
\label{log_hst}}
\tablehead{\colhead{Rootname} & \colhead{Filter} & \colhead{Exp. time}\\
          \colhead{}& \colhead{} &\colhead{(s)} }
\startdata
U2J20F01T  & F555W &  500   \nl 
U2J20F02T  &       &  500 \nl
U2J20F03T  &       &  500 \nl
U2J20F04T  &       &  160  \nl
U2J20F05T  & F814W &  400 \nl
U2J20F06T  &       &  400  \nl
\enddata
\end{deluxetable}

\begin{deluxetable}{lcc}
\tablecolumns{3}
\tablewidth{178 pt}
\tablecaption{Journal of Observations -- ESO spectra 
\label{log_eso}}
\tablehead{\colhead{Object} & \colhead{Pos. Angle} & \colhead{Exp. time}\\
          \colhead{}& \colhead{(deg.)} &\colhead{(s)} }
\startdata
NGC 3379  &  \phn70     & 2 \vzs\ 2700          \nl
          &  115          &  2 \vzs\ 2700        \nl
          &  160          &  2 \vzs\ 2700        \nl
\cutinhead{Kinematic standards}   \nl
HR 2305   &  par. angle     &  1  \nl   
HR 3110   &  par. angle     &  1 \nl
HR 25074   &  par. angle     &  1 \nl
HR 3122   &  par. angle     &  1 \nl
\cutinhead{Spectrophotometric standards} \nl
LTT 7379  &  par. angle     &   120 \nl
EG 274    &  par. angle     &  180  \nl
LTT 7987  &  par. angle     &  180  \nl
\enddata
\end{deluxetable}


\begin{deluxetable}{lcc}
\tablecolumns{3}
\tablewidth{215 pt}
\tablecaption{Photometric Parameters \label{phot_tab}}
\tablehead{\colhead{Parameter} & \colhead{V} & \colhead{I} }
\startdata
\cutinhead{Cuspy Law}
$I_0$  (mag) &  15.05\phn\ \mm\ 0.02\phn  &    12.98\phn\ \mm\ 0.02\phn \nl
$\gamma$     &  \phn0.210 \mm\ 0.004 &    \phn0.203 \mm\ 0.005 \nl
\cutinhead{S\'ersic Law}
$I_e$ (mag)   &  19.1\phn\ \mm\ 0.1\phn  &   17.0\phn\ \mm\ 0.2\phn \nl
$r_e$ (\arcsec) &  12.7\phn\ \mm\ 0.5\phn   &   12.4\phn\ \mm\ 0.6\phn  \nl
$n$             &  \phn2.36 \mm\ 0.05  &   \phn2.39 \mm\ 0.07 \nl 
\enddata
\end{deluxetable}


\begin{deluxetable}{lcc}
\tablecolumns{3}
\tablewidth{141 pt}
\tablecaption{Kinematical Parameters \label{starrot_tab}}
\tablehead{\colhead{Parameter} & \colhead{Model} }
\startdata
A (\kms)           & 113 (110 -- 117)  \nl
p	           &  1.0  \nl
C$_o$ (\arcsec)    & 5.8 (5.7 -- 5.9) \nl
$\Psi_o$ (\arcdeg) & 67 (67 -- 72) \nl
$i$ (\arcdeg)      & 26 (25 -- 28) \nl
\enddata
\end{deluxetable}



\clearpage

\begin{references}

\reference{bq90} Balcells, M. \& Quinn, P. J. 1990, \apj, 361, 381.

\reference{rbetal89} Bender, R., Capaccioli, M., Macchetto, F. D. \& Nieto, J.-L. 1989, The Messenger, 55, 6.

\reference{bsg94} Bender, R., Saglia, R. P. \& Gerhard, O. E. 1994, \mnras, 269, 785.

\reference{fbetal91} Bertola, F., Bettoni, D., Danziger, J., Sadler, E., Sparke, L. \& de Zeeuw, T. 1991, \apj, 373, 369.

\reference{jbetal96} Biretta, J. A., et al. 1996, WFPC2 Instrument Handbook, Version 4.0 (Baltimore:STScI).

\reference{c:m87} Capaccioli, M. 1987, in IAU Symp 127 ``Structure and Dynamics 
of Elliptical Galaxies'', ed. T. de Zeeuw (Dordrecht:Reidel), p. 47.

\reference{c:m90} Capaccioli, M. 1990, in ``Bulges of Galaxies'', ed. B. Jarvis \& D. M. Terndrup (Garching:ESO), p. 231.

\reference{mcetal90} Capaccioli, M., Held, E. V., Lorenz, H. \& Vietri, M. 1990, \aj, 99, 1813.

\reference{mcetal91} Capaccioli, M., Vietri, M., Held, E. V. \& Lorenz, H. 1991, \apj, 371, 535. (C91)

\reference{cjf89} Ciardullo, R., Jacoby, G. H. \& Ford, H. C. 1989, \apj, 344, 715.

\reference{cmw98} Cot\'e, P., Marzke, R. O. \& West, M. J. 1998, \apj, 501, 554.

\reference{pcetal93} Crane, P., et al. 1993, \aj, 106, 1371.

\reference{cdoetal91} Dalle Ore, C., Faber, S. M., Jesus, J., Stoughton, R. \& Burstein, D. 1991, \apj, 366, 38.

\reference{dvl88} de Vaucouleurs, A., \& Longo, G. 1988, ``Catalogue of Visual and Infrared Photometry of Galaxies from 0.5\mic\ to 10\mic\ (1963 -- 1985)'', Univ. of Texas Monographs in Astronomy (Austin:Univ. of Texas).

\reference{sfetal97} Faber, S. M., et al. 1997, \aj, 114, 1771.

\reference{ff99} Ferrarese, L. \& Ford, H. C. 1999, \apj, 515, 583. 

\reference{lfetal94} Ferrarese, L., van den Bosch, F. C., Ford, H. C., Jaffe, W. \& O'Connell, R. W. 1994, \aj, 108, 1598. 
 
\reference{ffetal99} Ferrari, F., Pastoriza, M. G., Macchetto, F. D. \& Caon, N. 1999, \aaps, 136, 269. 

\reference{ffi95} Forbes, D. A., Franx, M.  \& Illingworth, G. D. 1995, \aj, 109, 1988.

\reference{fih89} Franx, M., Illingworth, G. D., \& Heckman, T. 1989, \apj, 344, 613.

\reference{k:j99} Kormendy, J. 1999, in ``Galaxy Dynamics'', ed. D. R. Merrit, M. Valluri, \& J. A. Sellwood (San Francisco:ASP), in press.

\reference{kgetal96} Gebhardt, K., Richstone, D., Kormendy, J., Bender, R., Faber, S., Lauer, T., Magorrian, J., \& Tremaine, S. 1996, \baas, 189, 111.03.

\reference{hb91} Hernquist, L. \& Barnes, J. E. 1991, \nat, 354, 210. 
 
\reference{if89} Illingworth, G. D. \& Franx, M. 1989, in ``Dynamics of Dense 
Stellar Systems'', ed. D. Merrit (Cambridge:Cambridge Univ Press), p. 13.

\reference{l:t85} Lauer, T. R. 1985, \mnras, 216, 429.

\reference{tletal92} Lauer, T. R., et al. 1995, \aj, 103, 703.

\reference{tletal95} Lauer, T. R., et al. 1995, \aj, 110, 2622.

\reference{fdmetal96} Macchetto, F. D., Pastoriza, M. G., Caon, N., Sparks, W. B., Giavalisco, M., Bender, R. \& Capaccioli, M. 1996, \aaps, 120, 463. (Paper I)

\reference{jmetal98} Magorrian, J., et al. 1998, \aj, 115, 2285.

\reference{n:jl89} Nieto, J.-L. 1989, in ``Proc. 2nd Extragalactic Astronomy Regional Meeting'' (Cordoba, Argentina) (Cordoba:Academia de Ciencias), p. 239.

\reference{nbs91} Nieto, J.-L., Bender, R. \& Surma, P. 1991, \aap, 244, 25.

\reference{rfpetal90} Peletier, R. F., Davies, R. L., Davis, L. E., Illingworth, G. D. \& Cawson, M. 1990, \aj, 100, 1091.

\reference{pn94} Poulain, P. \& Nieto, J.-L. 1994, \aaps, 103, 573.

\reference{ssetal97} Sakai, S. , Madore, B. F., Freedman, W. L., Lauer, T. R., Ajhar, E. A. \& Baum, W. A. 1997, \apj, 478, 49. 

\reference{s:j68} S\'ersic, J. L. 1968, Atlas de Galaxias Australes (Cordoba:Universidad de Cordoba), p 141.
 
\reference{s:ts94} Statler, T. S. 1994, \aj, 108, 111.

\reference{ssh99} Statler, T. S. \& Smecker-Hane, T. 1999, \aj, 117, 839. (SS99)

\reference{sesetal76} Strom, S. E., Strom, K. M., Goad, J. W., Vrba, F. J. \& Rice, W. 1976, \apj, 204, 684.

\reference{vdbetal94} van den Bosch, F. C., Ferrarese, L., Jaffe, W., Ford, H. C. \& O'Conell, R. W., 1994, \aj, 108, 1579.

\reference{bjm98} van den Bosch, F. C., Jaffe, W. \& van den Marel, R. P. 1998, MNRAS, 293, 343.

\reference{vDf95} van Dokkum, P. G. \& Franx, M. 1995, \aj, 110, 2027.

\reference{vdm:rp91} van der Marel, R. P. 1991, \mnras, 2503, 710.

\reference{vdm:rp98} van der Marel, R. P. 1999, \aj, 117, 744.

\reference{y:p80} Young, P. 1980, \apj, 242, 1232.

\end{references}
\end{document}